\documentstyle[12pt]{article}
\begin{document}
\titlepage

\title{Regularization of superstring amplitudes and
a cancellation of divergences in superstring theory}
\author{G.S. Danilov\thanks{E-mail address:
danilov@thd.pnpi.spb.ru}\\
Petersburg Nuclear Physics Institute\\
Gatchina, St.-Petersburg 188350, Russia}
\date{}
\maketitle

\begin{abstract}
For a calculation of divergent fermion string amplitudes a
regularization procedure invariant under the supermodular group
is constructed. By this procedure superstring amplitudes of an
arbitrary genus are calculated using both partition functions and
superfield vacuum correlators computed early. A finiteness of
superstring amplitudes and related topics are discussed.
\end{abstract}

\newpage

\section{Introduction}
For a long time the central matter of multi-loop calculations in
superstring theory \cite{rnshw} was focused on partition
functions and of field vacuum correlators [2---9]. Essentially,
difficulties were found for Ramond strings where
the above values can not be derived by
an obvious extension of boson string results \cite{vec7}. A
method calculating the discussed values has been developed in
\cite{dan1,danphr}. By this method the partition functions and
the superfield vacuum correlators were calculated explicitly
\cite{danphr,dan6} in terms of super-Schottky group
\cite{dan3,danphr} parameters for all both the Ramond strings
and the Neveu-Schwarz ones. In
a calculation of superstring amplitudes through the
above values the main difficulty is due to every fermion string
amplitude is divergent. Removing the above difficulty is
considered in this paper.

As it is usually, a genus-$n$ closed
superstring amplitude $A_{n,m}$ with $m$ legs is given by
\begin{equation}
A_{n,m}=\int\prod_N dq_Nd\overline
q_N\prod_{r=1}^m dt_rd\overline t_r \sum_{L,L'}
Z_{L,L'}^{(n)}(\{q_N,\overline q_N\}) E_{L,L'}
(\{t_r,\overline t_r\})
\label{ampl}
\end{equation}
where the overline denotes the complex conjugation and $L$ ($L'$)
labels superspin structures of right (left) superfields
defined on the complex $(1|1)$ supermanifolds
\cite{bshw}. Every superspin structure $L=(l_1 ,l_2)$ presents a
superconformal extension of the
$(l_1,l_2)=\bigcup_{s=1}^n(l_{1s},l_{2s})$ ordinary spin one
\cite{swit}. The genus-$n$ theta function characteristics
$(l_1,l_2)$ can be restricted by $l_{is}\in(0,1/2)$. As the
$q_N$ moduli, $(3n-3|2n-2)$ of $(3n|2n)$ super-Schottky modular
parameters are used,  the rest being in a number of $(3|2)$, is
fixed commonly to all the genus-$n$ supermanifolds by a
super-M{\"o}bius transformation.  Partition functions
$Z_{L,L'}^{(n)}(\{q_N,\overline q_N\})$ are calculated from
equations \cite{dan1,danphr} expressing that the superstring
amplitudes are independent of a choice of two-dim.  metrics and
of a gravitino field. The vacuum expectation
$E_{L,L'}(\{t_r,\overline t_r\})$ of the vertex product depends
on the $\{t_r\}$ set of vertex supercoordinates and on
$\{q_N,\overline q_N\}$, as well. We map the supermanifolds by
the supercoordinate $t=(z|\theta)$ where z is a local complex
coordinate and $\theta$ is its odd partner. Every $t_r$
integration in (\ref{ampl}) is performed over the supermanifold.
Each of the $(L,L')$ terms is covariant \cite{dnph} under the
supermodular group, moduli being integrated over the fundamental
domain calculated in \cite{dnph}. Among other things, the above
fundamental domain depends on $L$ through terms proportional to
odd super-Schottky parameters because, generally, the
supermodular changes of moduli and of supercoordinates  depend
on superspin structure \cite{dnph}. For the same reason, the sum
over $(L,L')$ in (\ref{ampl}) calculated with $\{q_N\}$ common
to all the superspin structures is non-covariant under the group
considered.

The integration of every superspin structure in
(\ref{ampl}) is divergent \cite{belkniz} due to a degeneration
of Riemann surfaces. It is expected usually
\cite{ver,martpl,berk} that all the divergences are canceled in
the whole amplitude $A_{n,m}$,  but, in any case, a correct
calculation of $A_{n,m}$ requires a regularization procedure.
The regularization procedure is bound to ensure the invariance
of the superstring amplitudes under supermodular group.
At the same time, a cutoff  of modular integrals
proposed in \cite{grpd} violates the above group. Specifically,
in \cite{grpd} the non-split property of the supermanifolds
\cite{dnph} is ignored. There is no a test for
the supermodular invariance to be restored after removing the
cutoff \cite{grpd} once the integrals in (\ref{ampl}) are
calculated and the summation over superspin structures is
performed. In contrast, we build a manifestly supermodular
regularization of (\ref{ampl}) multiplying every term of the sum
by a supermodular invariant function. Besides the modular
integrals, we regularize integrals over $z_j$ that are ill
defined when all the vertices tend to coincide, or,
alternatively, all they are moved away from each other. As a
by-product, we construct a supermodular covariant sum over
superspin structures. A popular opinion is that all the
non-integrable singularities are cancelled locally in sum of
this kind, but it has not really a firm basis.
Indeed, the above local cancellation of the divergences requires
\cite{ver,martpl} vanishing the supercovariant sum over
superspin structures of the partition functions. It might be
reasonable \cite{marteo}, if this sum is essentially a product
of a function holomorphic in moduli by an anti-holomorphic one.
But except the genus-one case, such is not the case because the
supermanifold period matrices of a genus-$n>1$ depend
\cite{pst,danphr} on $L$. Further still, in this condition a
local nullification of the discussed sum of the partition
functions is not necessary for a vanishing \cite{martpl} of the
whole vacuum amplitude.  Moreover, even though \cite{berk}, it
is not established that divergences are locally cancelled
\cite{berk} in the multi-loop Green-Schwarz amplitudes because
in \cite{berk} the non-split property of the supermanifolds
\cite{dnph} is not taken into account. In our approach a
divergence cancellation in (\ref{ampl}) are studied once the
integrals are  calculated.

Poles and of threshold singularities of $A_{n,m}$ in a given
reaction channel $j$ are due to an integration over a suitable
nodal domain. As it is usual \cite{gsw}, each of the above
integrals is calculated at $Re\,E_j^2<0$ where $E_j$ is a center
mass energy in the channel considered.  Being divergent itself
at $Re\,E_j^2>0$, it is extended over $Re\,E_j^2>0$ by an
analytical continuation in $E_j^2$.  Since there is no a region
in the $\{E_j^2\}$ space where all the discussed integrals would
be finite together and since a boundaries of the nodal domains
are changed under the supermodular group, the supermodular
invariance of the considered procedure may seem doubtful. We
improve this procedure and clarify supermodular invariance of
it.

We expect that amplitudes of an
emission of a longitudinal polarized gauge boson vanish in our
scheme as it is required by the gauge symmetry. In addition, the
0-, 1-, 2- and 3- point functions of massless superstring modes
are nullified in line with \cite{martpl}.  A full study of the
above matters is planned in the next future.

The regularization procedure for the modular integrals is
considered in Sec.2. It is partly overlapped with \cite{dan7}
where a regularization of those was proposed. In Sec.3
regularized expressions for superstring amplitudes are given. In
Sec.4 the cancellation of divergences in superstring
amplitudes, the gauge symmetry  and non-renormalization theorems
\cite{martpl} are discussed.

\section{Regularization of modular integrals}

A construction of desired supermodular invariant
functions of moduli is complicated due to supermodular
changes of $q_N$ depend on the superspin structure \cite{dnph}.
To build the desired function we perform a singular
$t\rightarrow \hat t=(\hat z|\hat\theta)$ superholomorphic
transformation \cite{dan7} to a new parameterization $P_{split}$
where transition groups contain no Grassmann parameters:
\begin{equation}
z=f_L(\hat z)+f_L'(\hat
z)\hat\theta\xi_L(\hat z),\, \theta=\sqrt{f_L'(\hat
z)}\left[\left(1+\frac{1}{2}\xi_L(\hat z)\xi_L'(\hat z)\right)
\hat\theta+\xi_L(\hat z)\right]\,,\quad f_L(\hat
z)=\hat z+y_L(\hat z)\,.
\label{split}
\end{equation}
Here the "prime" symbolizes $\hat z$-derivative,
$\xi_L(\hat z)$ is a Grassmann function and the $y_L(\hat z)$
function is proportional to odd modular parameters.  Rounds
about $(A_s,B_s)$-cycles are given by super-Schottky
transformations $(\Gamma_{a,s}(l_{1s}),\Gamma_{b,s}(l_{2s}))$,
every $A_s$-cycle being associated with a suitable Schottky
circle.  In this case one see \cite{dan3,danphr} that
$\Gamma_{a,s}(l_{1s}=0)=I$, $\Gamma_{a,s}^2(l_{1s}=1/2)=I$, but
$\Gamma_{a,s}(l_{1s}=1/2)\neq I$. So a square root cut on the
z-plane appears for every $l_{1s}\neq0$ with endcut points to be
inside corresponding Schottky circles. In the
$P_{split}$ parameterization the same rounds are associated with
transformations
$(\hat\Gamma_{a,s}(l_{1s}),\hat\Gamma_{b,s}(l_{2s}))$. In this
case
\begin{equation}
\Gamma_{b,s}(l_{2s})(t)=
t\left(\hat\Gamma_{b,s}(l_{2s})(\hat t)\right)\,,\quad
\Gamma_{a,s}(l_{1s})(t)=
t^{(s)}\left(\hat\Gamma_{a,s}(l_{1s})(\hat t)\right)
\label{main}
\end{equation}
where $t^{(s)}(\hat t)$ is obtained by $2\pi$-twist of $t(\hat
t)$ on the complex $\hat z$-plane about the Schottky circle
assigned to a particular handle $s$.
Both $\hat\Gamma_{a,s}(l_{1s})$ and
$\hat\Gamma_{b,s}(l_{2s})$ do not contain
Grassmann  modular parameters. Since the super-Schottky
transformations depend, among other things, on $(2n-2)$
Grassmann moduli, the transition functions in (\ref{split})
depends on $(2n-2)$ Grassmann parameters
$(\lambda_j^{(1)},\lambda_j^{(2)})$ where $j=1\dots n-1$.
Eqs.(\ref{main}) are given explicitly in \cite{dan7}. Kindred
equations were used in \cite{dnph} to calculate the acting of
supermodular transformations on supercoordinates and on modular
parameters. Unlike those in \cite{dnph}, eqs.(\ref{main}) can be
satisfied only if the transition functions in (\ref{split}) have
poles in a fundamental region of $\hat z$-plane. We take
\footnote{An another choice of the poles is discussed in
\cite{dan7}.} them possessing $(n-1)$ poles $\hat z_j$ of an
order 2. For even superspin structures we choose the above poles
among $n$ zeros of the fermion Green function $R_L^f(\hat z,\hat
z_0)$ calculated for zero odd moduli.\footnote{See Sec. 4 of
\cite{danphr} where $R_L^f(\hat z,\hat z_0)$ is denoted as
$R_f(z,z')$.}  For odd superspin structures the poles can be
chosen by a similar way \cite{dan7}. We take $\hat z_0$ common
to all superspin structures. In this case supermodular changes
of $(\lambda_j^{(1)},\lambda_j^{(2)})$ are independent of the
superspin structure and the supermodular group in the
$P_{split}$ representation is mainly reduced to the ordinary
modular one.  The singular parts of (\ref{split}) are determined
by a condition that above modular group is isomorphic to the
supermodular one in the super-Schottky parameterization. In this
case one obtains \cite{dan7} near every pole $\hat z_j(\hat
z_0;L)$ that
\begin{eqnarray}
\xi_L(\hat z) \approx\frac{[1+ \xi_L(\hat
z)\xi_L'(\hat z)]}{R_L^f(\hat z,\hat z_0)} \left[
\frac{\lambda_j^{(2)} }{[R_L^f(\hat z,\hat z_0)]}\frac{\partial
R_L^f(\hat z,\hat z_0)}{\partial_{\hat z_0}}
+\lambda_j^{(1)}\right]
+\frac{\lambda_j^{(1)}\lambda_j^{(2)}\xi_L(\hat z)}
{2[R_L^f(\hat z,\hat z_0)]^2}
\frac{\partial^2\ln[R_L^f(\hat z,\hat z_0)]}{\partial_{\hat
z}\partial_{\hat z_0}} \,, \nonumber\\
f_L(\hat z)\approx
\frac{\lambda_j^{(2)}
\xi_L(\hat z)f_L'(\hat z)}{[R_L^f(\hat z,\hat z_0)]^2}
\frac{\partial
R_L^f(\hat z,\hat z_0)}{\partial_{\hat z_0}}
+\frac{\lambda_j^{(1)}
\xi_L(\hat z)f_L'(\hat z)}{R_L^f(\hat z,\hat z_0)}
\label{ppart}
\end{eqnarray}
where the "prime" symbol denotes $\partial_{\hat z}$. The
calculation of both $y_L(\hat z)$ and $\xi_L(\hat z)$ is quite
similar to that in Sec. 3 of \cite{dnph}. For this purpose we
start with relations
\begin{equation}
\xi_L(\hat z)=-\int_{C(\hat z)}
G_{gh}^{(f)}(\hat z,\hat z';L)\xi_L(\hat z')
\frac{d\hat z'}{2\pi i}\,, \quad
y_L(\hat z)=\int_{C(\hat z)}G_{gh}^{(b)}(\hat z,\hat z')
y_L(\hat z')
\frac{d\hat z'}{2\pi i}
\label{inf}
\end{equation}
where an infinitesimal contour $C(z)$ gets around $z$-point in
the positive direction. Both Green functions $G_{gh}^{(f)}(\hat
z,\hat z';L)$ and $G_{gh}^{(b)}(\hat z,\hat z')$ are
defined\footnote{$G_{gh}^{(f)}(\hat z,\hat z';L)$ is denoted in
\cite{dnph} as $G_{gh}^{(f)}(\hat z,\hat z')$, the explicit $L$
dependence being omitted.} in \cite{dnph}. We deform the $C(\hat
z)$ to surround both the $\hat z_j$ poles and the Schottky
circles together with the cuts presenting for $l_{1s}\neq0$.
The integrals along Schottky circles and along the cuts are
transformed by (\ref{main}) to the form similar to \cite{dnph}.
The integrals around the poles are calculated using
(\ref{ppart}). Relations for modular parameters appear
just as in Sec.4 of \cite{dnph} because of $G_{gh}^{(f)}(\hat
z,\hat z';L)$ and $G_{gh}^{(b)}(\hat z,\hat z')$ receive
additional terms under Schottky transformations on $\hat
z$-plane.\footnote{See eqs.(23) in \cite{dnph}.} The resulted
equations \cite{dan7} determine both
$y_L(\hat z)$ and $\xi_L(\hat z)$ together with complex
super-Schottky moduli $q_N$ up to $SL_2$ transformations of $t$.
The partition functions $\hat Z_{L,L'}^{(n)}
(\{\hat q_N,\overline{\hat
q_N}\})$ in the $P_{split}$ representation are given by
\begin{equation}
\hat Z_{L,L'}^{(n)}(\{\hat q_N,\overline{\hat q_N}\})=
F_L(\{\hat q_N\})\overline{F_{L'}(\{\hat q_N\})}
Z_{L,L'}^{(n)}(\{q_N,\overline q_N\})
\label{zspl}
\end{equation}
where $F_L(\{\hat q_N\})$ is Jacobian of the considered
transformation and $q_N=q_N(\{\hat q_N\};L)$. Further,
$\{\hat q_N\}=\{\hat q_{ev},\lambda_j^{(1)},\lambda_j^{(2)}\}$,
where $\hat q_{ev}$ are even complex moduli in a number of
$(3n-3)$.  The supermodular invariant sum over superspin
structures is constructed by doing
the $(\{\hat q_N\},\hat t)$ set to be
common to all superspin structures. Superstrings are
non-invariant under the considered singular transformations
(\ref{split}). That is why expressions for the amplitudes in the
$P_{split}$ parameterization  (once integrations over
$\{\lambda_j^{(1)},\lambda_j^{(2)}\}$ are performed) differ from
those in \cite{ver} where a split property of the
supermanifolds is assumed (for more details, see \cite{dan7}).

The desired supermodular invariant function
$Y(\{\hat q_N,\overline{\hat q_N}\};\hat z_0,\overline{\hat
z_0})$ is constructed by (\ref{zspl}) as
\begin{equation}
Y(\{\hat
q_N,\overline{\hat q_N}\};\hat z_0,\overline{\hat z_0})=\frac{
[Y_1(\{\hat q_N,\overline{\hat q_N}\};\hat z_0,\overline{\hat
z_0})]^{2^{n-1}(2^n+1)}} {Y_2(\{\hat q_N,\overline{\hat
q_N}\};\hat z_0,\overline{\hat z_0})}
\label{reg}
\end{equation}
with $Y_1(\{\hat q_N,\overline{\hat q_N}\};\hat
z_0,\overline{\hat z_0})\equiv Y_1$ and
$Y_2(\{\hat q_N,\overline{\hat q_N}\};\hat
z_0,\overline{\hat z_0})\equiv Y_2$ defined to be
\begin{equation}
Y_1=
\sum_{L\in\{L_{ev}\}}
\hat Z_{L,L}^{(n)}(\{\hat q_N,\overline{\hat
q_N}\})\quad{\rm and}\quad
Y_2(\{\hat q_N,\overline{\hat q_N}\};\hat
z_0,\overline{\hat z_0})=
\prod_{L\in\{L_{ev}\}}\hat
Z_{L,L}^{(n)}(\{\hat q_N,\overline{\hat q_N}\})
\label{regul}
\end{equation}
where $\{L_{ev}\}$ is the set of $2^{n-1}(2^n+1)$ even spin
structures and $q_N=q_N(\{\hat q_N\};L)$. Since both $Y_1(\{\hat
q_N,\overline{\hat q_N}\})$ and $Y_2(\{\hat q_N,\overline{\hat
q_N}\})$ receive the same factor under modular transformation of
$\hat q_N$-parameters, the right side of (\ref{reg}) is
invariant under supermodular transformations. In addition, it
tends to infinity, if Riemann surfaces are degenerated. Indeed,
if a particular handle, say $s$, become degenerated, the
corresponding Schottky multiplier $k_s$ tends to zero. In this
case both the nominator and the denominator in (\ref{reg}) tend
to infinity \cite{danphr}, but terms associated with $l_{1s}=0$
have an additional factor $|k_s|^{-1}\rightarrow\infty$ in a
comparison with those associated with non-zero $l_{1s}$. So
$Y(\{\hat q_N,\overline{\hat q_N}\};\hat z_0,\overline{\hat
z_0})\rightarrow\infty$. If a even spin structure of a
genus-$n>1$ is degenerated into odd spin structures, the
partition functions tend to zero \cite{dan6} and they do not
vanish, if it is degenerated into even spin ones. So again
$Y(\{\hat q_N,\overline{\hat q_N}\};\hat z_0,\overline{\hat
z_0})\rightarrow\infty$. Hence to regularize the desired
integrals one can introduce in the integrand (\ref{ampl}) a
multiplier
\begin{equation}
B_{mod}^{(n)}(\{q_N,\overline{q_N}\};\hat z_0,
\overline{\hat z_0};\delta_0)=\{\exp[-\delta_0
Y(\{\hat q_N,\overline{\hat q_N}\};\hat z_0,\overline{\hat
z_0})]\}_{sym}
\label{modreg}
\end{equation}
where $\delta_0>0$ is a parameter. The right side of
(\ref{modreg}) is symmetrized over all the sets of $(n-1)$ zeros
of the fermion Green function $R_L^f(\hat z,\hat z_0)$.
By the above reasons, (\ref{modreg}) vanishes, if Riemann
surfaces become degenerated that provides the
finiteness of the modular integrals in (\ref{ampl}).  It is
follows from (\ref{main}) and (\ref{ppart}) that (\ref{modreg})
is invariant under $L_2$-transformations of $\hat z_0$
accompanied by suitable transformations of both
$\lambda_j^{(1)}$ and $\lambda_j^{(2)}$.  For a given $\hat
t_0=(\hat z_0|\hat\theta=0)$ one can calculate its image $\tilde
t=(z_0|\tilde\theta(z_0))$ under the mapping (\ref{split}). In
this case the transition functions have no poles because zeros
$\hat z_j(\hat z_0;L)$ of $R_L^f(\hat z,\hat z_0)$ are always
different from $\hat z_0$. Simultaneously, so far as $\hat
z_j(\hat z_0;L)$ is changed under fundamental group
transformations, eqs.(\ref{main}) are satisfied only if every
this transformation is accompanied by an appropriate change of
the $(\lambda_j^{(1)},\lambda_j^{(2)})$ parameters that is
calculated from (\ref{ppart}).  Because the above change of
$(\lambda_j^{(1)},\lambda_j^{(2)})$ does not depend on the
superspin structure, (\ref{modreg}) is invariant under the
super-Schottky transformations of $\tilde t$.

\section{Superstring amplitudes}

To regularize the integrals over $t_j$, we need functions
depending on two more supermanifold points $t_a=(t_{-1},t_0)$ in
addition to $\{t_j\}$.  One receives in hands the above $t_a$
points multiplying (\ref{ampl}) by the unity arranged to be a
square in the same integrals, every integral
$I_{LL'}^{(n)}=1$ being
\begin{eqnarray}
I_{LL'}^{(n)}=\frac{1}{n}\int\frac{dtd\overline
t}{2\pi i} I_{LL'}^{(n)}(t,\overline t)\,,\quad I_{LL'}^{(n)}
(t,\overline t)= D(t)[J_s(t;L)+\overline{J_s(t;L')}]
[2\pi i\omega(L)-2\pi
i\overline{\omega(L')}]_{sr}^{-1}\nonumber\\ \times
\overline{D(t)}[J_r(t;L)+\overline{J_r(t;L')}]\,,\quad
D(t)=\theta\partial_z+\partial_{\theta}\,.
\label{illin}
\end{eqnarray}
Here $J_s(t;L)$ are the genus-$n$ superholomorphic functions
\cite{danphr} having periods, $\omega_{sr}(L)$ is a period
matrix on the supermanifold and $D(t)$ is the spinor derivative.
The above period matrix depends on the superspin structure
\cite{pst,danphr}. Owing to $\overline{D(t)}J_r(t;L)=0$, both
$\overline{J_s(t;L')}$ and $J_r(t;L)$ can be omitted in
(\ref{illin}), but we remain them to have the integrand without
cuts on the supermanifold.  Integrating (\ref{illin}) by parts
one obtains that $I_{LL'}^{(n)}=1$ as it was announced. With
(\ref{illin}), we define a regularized superstring amplitude
$A_{n,m}(\{\delta\})$ with $m>3$ as
\begin{eqnarray}
A_{n,m}(\{\delta\})=\int\left(
\prod_N dq_Nd\overline q_N\right)\left(\prod_{r=1}^m
dt_rd\overline t_r\right)\sum_{L,L'}
\left(\prod_{a=-1}^0dt_ad\overline t_a
I_{LL'}^{(n)}(t_a,\overline t_a)\right)
Z_{L,L'}^{(n)}(\{q_N,\overline q_N\})
\nonumber \\ \times
E_{L,L'}(\{t_r,\overline t_r\})
B_{mod}^{(n)}(\{q_N,\overline{q_N}\};\hat z_0,\overline{\hat
z_0};\delta_0)
\prod_{(jl)}
B_{jl}^{(n)}(\{t_a,\overline t_a\};\{q_N,\overline{q_N}\};
\{\delta_{jl}\};L,L')
\label{reampl}
\end{eqnarray}
where $t=(z_0|\theta)$, the $(jl)$ symbol labels pairs of the
vertices, $\delta_{jl}>0$ are parameters and
$\{\delta\}=(\delta_0,\{\delta_{jl}\})$. Further, $\hat z_0=\hat
z_0(z_0)$ is calculated together with $\theta(z_0)$ from
(\ref{split}) taken at both $\hat\theta=0$, $z=z_0$ and
$\theta=\theta(z_0)$. At $\{\delta_{jl}>0\}$ every
$B_{jl}^{(n)}(\{t_a,\overline t_a\};\{q_N,\overline{q_N}\};
\{\delta_{jl}\};L,L')$ factor tends to zero at
$|z_j-z_l|\rightarrow0$ and at $|z_j-z_l|\rightarrow\infty$. The
integration domain over moduli has been calculated in
\cite{dnph}. Also, one can rewrite (\ref{reampl}) in the
$P_{split}$ variables. In this case moduli are integrated over
the fundamental domain of the ordinary modular group. In details
Grassmann integrations in (\ref{reampl}) are planned to discuss
elsewhere. The superstring amplitude $A_{n,m}$ is defined as
$A_{n,m}(\{\delta\rightarrow0\})$ calculated in line with the
usual analytical continuation procedure \cite{gsw} for the
integrals over nodal domains giving rise to poles and threshold
singularities of $A_{n,m}$. The regularization factors
$B_{jl}^{(n)}(\{t_a,\overline t_a\};\{q_N,\overline{q_N}\};
\{\delta_{jl}\};L,L')$ are calculated in terms
of the supermodular scalars $U_{jl}(\{t_a,\overline
t_a\};\{q_N,\overline q_N\};L,L')$ defined by
\begin{equation}
U_{jl}(\{t_a,\overline t_a\};\{q_N,\overline q_N\};L,L')=
\exp\left[2X_{j,l}+2X_{-1,0}-X_{j,-1}-X_{j,0}-
X_{l,-1}-X_{l,0}\right]
\label{ufunct}
\end{equation}
where $X_{r,s}\equiv
X_{L,L'}(t_r,\overline{t_r};t_s,\overline{t_s})$ are the
vacuum correlators of the scalar superfields. With
(\ref{ufunct}), the desired factors in (\ref{reampl})
can be constructed as
\begin{equation}
B_{jl}^{(n)}(\{t_a,\overline t_a\};\{q_N,\overline{q_N}\};
\{\delta_{jl}\};L,L')=
\left[\frac{ U_{jl}}{1+
U_{jl}^{2}}\right]^{\delta_{jl}^0}
\exp[-\delta_{jl}^{(1)}U_{jl}-
\delta_{jl}^{(2)}U_{jl}^{-1}]
\label{bjl}
\end{equation}
where $U_{jl}\equiv
U_{jl}(\{t_a,\overline t_a\};\{q_N,\overline q_N\};L,L')$ is
given by (\ref{ufunct}). Grassmann integrations are well defined
only for super-functions bounded together with all derivatives
thereof \cite{leites} that is just provided by the exponential
factor in (\ref{bjl}). For $(p_j+p_l)^2<0$ we take
$\delta_{jl}^0=0$. We define the scalar product of 10-momenta by
$p_jp_l=p_j^0p_l^0-p_j^np_l^n$. So the factor in front of the
exponential sign in (\ref{bjl}) presents only, if
$p_{jl}^2=(p_j+p_l)^2\geq0$. Without the discussed factor nodal
domain contributions to (\ref{reampl}) that originate
singularities of $A_{n,m}$ in $p_{jl}^2$-variables would tend to
infinity at $\{\delta\rightarrow0\}$ because of terms
$\exp[-p_{jl}^2\ln\tilde\delta_{jl}]$ with
$\tilde\delta_{jl}\in (\delta_{jl}^{(1)},\delta_{jl}^{(2)})$ or
$\tilde\delta_{jl}=\delta_0$. The factor in question presents,
$p_{jl}^2$ is replaced by
$p_{jl}^2-\delta_{jl}^0$. So the discussed
nodal domain integrations remain finite at
$\{\tilde\delta_{jl}\rightarrow0\}$, if
$Re\,p_{jl}^2<\delta_{jl}^0$. In this case the superstring
amplitude $A_{n,m}$ are obtained from (\ref{reampl}) by the
following manifestly supermodular invariant procedure.

One calculates $A_{n,m}(\{\delta\})$ at
$(\{\delta_{jl}^{(1)}=0,\delta_{jl}^{(2)}=0\}, \delta_0=0)$
in the domain where $Re\,p_{jl}^2<\delta_{jl}^0$ for every
$p_{jl}^2$. If divergences are really cancelled in $A_{n,m}$,
the result is finite due to (\ref{bjl}). Next, for a particular
$p_{jl}^2$ with $Re\,p_{jl}^2>0$, one performs an analytical
continuation to $Re\,p_{jl}^2<0$. Thereafter $\delta_{jl}^0$ is
taken be zero. This procedure is performed step by step for
every $p_{rs}^2$ with $Re\,p_{rs}^2>0$, the $A_{n,m}$ amplitude
is obtained in a certain region of the $\{p_{jl}^2\}$ space.

\section{ Divergence cancellation and related topics}

The procedure at the end of the previous Section verifies the
supermodular invariance of $A_{n,m}$. In action all the
$\delta_{jl}$ parameters can be nullify in (\ref{reampl}) after
a suitable rewriting  of the integrals over the nodal
domains following an appropriate super-M{\"o}bius transformation
of $\{t_j\}$ (as it will be reported elsewhere). Really the
factors (\ref{bjl}) are important in (\ref{reampl}) essentially
to verify that the above rearrangement of (\ref{reampl}) does
not violate the supermodular group. Hence we expect that the
gauge symmetry inherent to massless modes presents in $A_{n,m}$
though in (\ref{reampl}) it is violated due to the discussed
factors (\ref{bjl}). A detailed study of the matter is in
progress.

The 0-, 1-, 2- and 3- point functions do not depend on
of 10-momenta of interacting particles. They are calculated
from the factorization requirement on $A_{n,m}$ with
$m>3$ when a cluster $\{N_1\}$ of handles in a number of
$n_1<n$ is separated from the remainder $\{N_2\}$ of ones so
that $d/l\rightarrow0$. Here $d$ is a size of the cluster and
$l$ is a typical distance between it and the remaining handles.
In particular, 0-, 1-, 2- and 3-point functions of massless
superstring modes can be defined to be zeros, if it is
consistent with the above factorization requirement. By the
grounds of the preceding paragraph one obtains from
(\ref{reampl}) that the discussed massless
functions contribute to $A_{n,m}$ proportionally to either
$A_{n_1,0}^{(n)}$ or $n_1A_{n_1,1}^{(n)}$ to be
\begin{eqnarray}
A_{n_1,1}^{(n)}=\int \left(\prod_{N_1}
dq_{N_1}d\overline q_{N_1}\right)dtd\overline t\sum_{L_1,L_1'}
Z_{L_1,L_1'}^{(n_1)}(\{q_{N_1},\overline q_{N_1}\})
B_{mod}^{(n)}(\{q_N,\overline{q_N}\};\hat z_0,\overline{\hat
z_0};\delta_0)I_{L_1L_1'}^{(n_1)}(t,\overline t) \,,
\nonumber\\
A_{n_1,0}^{(n)}=\int \left(\prod_{N_1}
dq_{N_1}d\overline q_{N_1}\right)\sum_{L_1,L_1'}
Z_{L_1,L_1'}^{(n_1)}(\{q_{N_1},\overline q_{N_1}\})
B_{mod}^{(n)}(\{q_N,\overline{q_N}\};\hat z_1,\overline{\hat
z_0};\delta_0)
\label{regvac}
\end{eqnarray}
where $\hat z_1$ a point distanced from the $\{N_1\}$ cluster in
the order of $l$, superspin structures in the above cluster are
labeled by $(L_1,L_1')$ and $I_{L_1L_1'}^{(n_1)}(t,\overline t)$
is defined in (\ref{illin}) at $n=n_1$. Further,
$t=(z_0|\theta)$ and $\hat z_0=\hat z(z_0)$ is calculated just
as in (\ref{reampl}). Through the regularization factor
(\ref{modreg}) the integrands in (\ref{regvac}) depend on the
variables associated with the remainder $\{N_2\}$ of the handles
as well as on the $\{N_1\}$ cluster ones. The discussed
functions of massless superstring modes are nullified as it is
requested \cite{martpl}, if the right sides of
eqs.(\ref{regvac}) vanish at
$(d/l\rightarrow0,\delta_0\rightarrow0,d/l\gg\delta_0)$.
A preliminary study gives evidence that this
is really in the case as it is reported below.

Calculating (\ref{regvac}) under conditions of interest we
specify the $\{N_0\}$ set  of the $(3|2)$
super-Schottky parameters that are not moduli, say, as
$N_0=(u_1,v_1,u_2|\mu_1=0,\nu_1=0)$ with a notation
$(u_s|\mu_s)$ and $(v_s|\nu_s)$ for supercoordinates of two
fixed points of the super-Schottky transformation associated
with a given handle $s$. The right sides of (\ref{regvac}) do not
depend on $(u_1,v_1)$, as well as the right side of
(\ref{reampl}). Except only the $A_{1,0}^{(n)}$ case,
the leading term of the exponent in
(\ref{modreg}) is invariant under a super-M{\"o}bius
transformation of the super-Schottky fixed points assigned to
the $\{N_1\}$ cluster. If the handle $s=1$ does not belong to
the above cluster, this transformation removes in (\ref{regvac})
a dependence on two Grassmann variables. So both the integrals
(\ref{regvac}) are nullified owing to Grassmann integrations. In
the $A_{1,0}^{(n)}$ case the leading term of the exponent in
(\ref{modreg}) does not depend on a spin structure assigned to
$s=1$. Hence $A_{1,0}^{(n)}$ vanishes due to a nullifying of the
sum of the genus-1 partition functions. If $s=1$
belongs to the $\{N_1\}$ cluster, one can simplify the right
sides of (\ref{regvac}) taking $|u_1-v_1|\rightarrow0$ with
$|u_1-v_1|\gg\delta_0\rightarrow0$. In so doing the leading
term  in $A_{1,0}^{(n)}$ and in $A_{1,1}^{(n)}$ is proportional
to $A_{1,0}^{(n)}$ because the $s=1$ handle is separated from
the rest of the $N_1$ cluster and from $\hat z_0$. Hence in this
case the desired values disappear due to a vanishing of
$A_{1,0}^{(n)}$. If the discussed values (\ref{regvac}) vanish, a
degeneration of the genus-$n$ surfaces into a few ones of lower
genus does not originated divergences in the $A_{n,M}$
amplitudes of interest. In addition, the divergences due to a
degeneration of a particular handle are cancelled as it is
usually \cite{vec8,dnph}. So the superstring amplitudes appears
free from divergences. What values (\ref{regvac}) vanish it was
concluded from an examination only a leading term of the
exponent in (\ref{modreg}). An estimation of corrections
it is yet necessary, especially, in regions where a number of
zeros of the $R_L^f(\hat z,\hat z_0)$ fermion Green functions
coincide. A detailed calculation of (\ref{regvac}) is in
progress.

The work is supported by grant No. 96-02-18021 from the Russian
Fundamental Research Foundation.

\end{document}